# Characteristics of Interplanetary Discontinuities in the Inner Heliosphere Revealed by Parker Solar Probe

Y. Y. Liu[1,2], H. S. Fu[1,2,†], J. B. Cao[1,2], C. M. Liu[1,2], Z. Wang[1,2], Z. Z. Guo[1,2],

Y. Xu[1,2], S. D. Bale[3], and J. C. Kasper[4]

[1]School of Space and Environment, Beihang University, Beijing, 100191, China

[2]Key Laboratory of Space Environment Monitoring and Information Processing, Ministry of Industry and Information Technology, Beijing, 100191, China

[3]Space Sciences Laboratory and Physics Department, University of California, Berkeley, CA 94720-7450, USA

[4]Department of Climate and Space Sciences and Engineering, University of Michigan, Ann Arbor, MI 48109, USA.

†Correspondence to H. S. Fu at huishanf@gmail.com

**Abstract.**

We present a statistical analysis for the characteristics and spatial evolution of the interplanetary discontinuities (IDs) in the solar wind, from 0.13 to 0.9 au, by using the Parker Solar Probe measurements on Orbits 4 and 5. 3948 IDs have been collected, including 2511 rotational discontinuities (RDs) and 557 tangential discontinuities (TDs), with the remnant unidentified. The statistical results show that (1) the ID occurrence rate decreases from 200 events/day at 0.13 au to 1 events/day at 0.9 au, following a spatial scaling $r^{-2.00}$, (2) the RD to TD ratio decreases quickly with the heliocentric distance, from 8 at r<0.3 au to 1 at r>0.4 au, (3) the magnetic field tends to rotate across the IDs, 45° for TDs and 30° for RDs in the pristine solar wind within 0.3 au, (4) a special subgroup of RDs exist within 0.3 au, characterized by small field rotation angles and parallel or antiparallel propagations to the background magnetic fields, (5) the TD thicknesses normalized by local ion inertial lengths ($d_i$) show no clear spatial scaling and generally range from 5 to 35 $d_i$, and the normalized RD thicknesses follow $r^{-1.09}$ spatial scaling, (6) the outward (anti-sunward) propagating RDs predominate in all RDs, with the propagation speeds in the plasma rest frame proportional to $r^{-1.03}$. This work could improve our understandings for the ID characteristics and evolutions and shed light on the study of the turbulent environment in the pristine solar wind.

# 1. Introduction

Interplanetary discontinuities (IDs) are abundant in the solar wind, characterized by abrupt changes in the vector magnetic fields and/or the plasma properties (Ness et al. 1966; Lepping & Behannon 1986; Tsurutani et al. 1979, 1996; Neugebauer et al. 1984, 2006, 2010; Sonnerup et al. 2006, 2010; Horbury et al. 2001; Zhang et al. 2008a; Pallocchia et al. 2010; Paschmann et al. 2013). The Rankine-Hugoniot (R-H) condition predicts the existence of four types of IDs, including the shocks and contact/rotational/tangential discontinuities. The rotational and tangential discontinuities (RDs & TDs) are the most common types observed in the solar wind (Smith 1973; Mariani et al 1983; Paschmann et al. 2013). TDs can be treated as stationary boundaries in the plasma rest frame separating distinct plasmas (Burlaga & Ness 1969; Liu et al. 2019; Fu et al. 2012). RDs, however, are propagating kinks in the magnetic and flow fields (Smith 1973). The jumps of the magnetic field and plasma velocity across the RDs satisfy the Walén relation (Sonnerup et al. 1987). Theoretically, the RDs do not affect the plasma properties except for the velocity, with plasma density, entropy, and temperature remaining steady across them.

The IDs are believed to be closely related to the Alfvénic turbulence and generated locally because the observed log-normal distribution of discontinuity separations is consistent with the multiplicative random cascades and intermittency in 2D/3D simulations of MHD turbulence (Vasquez et al. 2007; Greco et al. 2008, 2009; Borovsky 2010; Servidio et al. 2011; Zhdankin et al. 2012; Yang et al. 2015, 2017). Some IDs, observed individually with large rotation of the magnetic field, appear usually as the boundaries of the magnetic field structures in the solar wind, such as the flux tubes, magnetic switchbacks, interplanetary coronal mass ejection and so on (Borovsky 2008; Zhang et al. 2008b; Mozer et al. 2020; Krasnoselskikh et al. 2020). The phase-steepened Alfvén waves (Tsurutani et al. 2002a, 2002b, 2005) are another possible source of the IDs in space.

These IDs are host to many dynamic processes, including magnetic reconnection, wave-particle interaction and Fermi acceleration (Gosling et al. 2007, 2009; Servidio et al. 2011; Huang et al. 2015; Liu et al. 2018a, 2018b, 2019; Fu et al 2020), and thus their characteristics and evolution have attracted the most interest. Much effort has been devoted into revealing the ID occurrence rate, the proportions of different types of IDs, the jump conditions, orientations, thicknesses, and the spatial scaling of these parameters (Neugebauer et al, 1984, 2010; Söding et al. 2001; Horbury et al. 2001; Tsurutani et al. 1979, 1996; Lepping & Behannon 1986). However, diverse or even conflicting views, were sometimes proposed on these issues based on the in-situ observations of the solar wind at different distances. Such disparities are mostly attributed to different identification criteria of the IDs, different survey regions, or different mechanisms at work predominantly affecting the ID properties.

The heliosphere within 0.3 au, exhibiting the greatest gradients of the solar wind properties (e.g., the magnetic field magnitude, the plasma density) in the radial direction, is the ideal natural laboratory to investigate the spatial evolution of the IDs but never in situ explored until the recent launch of Parker Solar Probe mission (PSP). In this paper, we perform a statistical analysis for the characteristics and spatial evolution of the IDs by using PSP data in the solar wind from 0.13 to 0.9 au.

## 2. Data Analysis

The Parker Solar Probe mission, launched on 2018 August 12 and carrying four instrument suites, has collected vast quantities of in-situ data of the pristine solar wind, in a previously unexplored region as close to the Sun as 0.13 au. The Fields Experiment (FIELDS; Bale et al. 2016) instrument suite on board PSP consisting of five voltage probes, two fluxgate magnetometers and one search coil magnetometer, is designed to measure the electric and magnetic fields from DC range to beyond the electron plasma frequency. The Solar Wind Electrons, Alphas, and Protons instrument suite (SWEAP; Kasper et al. 2016) contains the Solar Probe Cup (SPC; Case et al. 2020), a Sun-pointed Faraday Cup designed primarily to measure the proton moments, and the Solar Probe Analyzers (SPANs), a combination of two electrostatic analyzers on the ram side (for protons and electrons, respectively) and one on the anti-ram side (for electron only) for the measurement of 3D velocity distribution. In this study, we utilize the DC magnetic field data from FIELDS as well as the proton density and velocity data from SPC derived through the "moment" algorithm, on the PSP Orbits 4 (from December 1, 2019 to April 3, 2020) and 5 (from April 3, 2020 to August 1, 2020). The design sampling frequencies of the FIELDS instrument range between 2.3 to 293 Hz (Bale et al. 2019). In fact, such frequencies mostly exceed 9 Hz even under Cruise mode during Orbits 4 and 5, which is helpful for ID detection. For the convenience of data processing, the original magnetic field data are linearly interpolated into a uniform frequency of 25 Hz.

An automated selection algorithm is applied to the field data to find ID events, as described below. The selection criterion used here should meet the requirements that (1) the criterion should be adapted to the length scale of the field structures in the innermost heliosphere near 0.13 au, where the local proton gyroradii are an order of magnitude smaller than the typical values at 1 au, (2) the criterion should not be biased towards any specific type of discontinuities, and (3) the criterion should be able to distinguish the discontinuities from the stochastic magnetic field fluctuations. Thus, an ID in our selection algorithm is defined as the abrupt changes of the magnetic field, satisfying $|\Delta \mathbf{B}| > \frac{1}{3}|\mathbf{B}|_{mean} + 5$, where $|\Delta \mathbf{B}|$ and $|\mathbf{B}|_{mean}$ represent the field jump in two seconds and the mean magnetic field magnitude in a 2-second window corresponding to $|\Delta \mathbf{B}|$, respectively. The constant term 5nT is set on the right side to eliminate the effect from the stochastic field fluctuations in weak field region that otherwise are likely to be identified as IDs. With such a criterion, we scour the eight-month field data via automated algorithm to establish an ID event list. Then the minimum variance analysis (MVA; Mazelle et al. 1997; Cao et al. 2013) is performed to each ID to find the local magnetic normal (LMN) coordinate system, in which L corresponds to the maximum variance component, N coincides with the normal of the discontinuity plane (i.e., the direction along which the component of the magnetic vector data exhibit the smallest variance), and M completes the right-hand coordinate system. To ensure the accuracy of MVA results, only when the ratio of the middle to the minimum eigenvalue is larger than 2, are the results considered reliable and used for further analysis.

There are a diversity of specific classification criteria for TDs and RDs (Burlaga 1969; Tsurutani & Smith 1979; Neugebauer & Giacalone 2010), though the mathematical definitions of these IDs are uniform within the magnetohydrodynamics framework. Here we employ the criteria used by *Smith* (1973) with an additional modification to make it suitable for near-sun environment. An ID is identified as a TD when $\frac{|B_N|}{|\mathbf{B}|_{mean}} < 0.2$ or otherwise a RD when $\frac{|B_N|}{|\mathbf{B}|_{mean}} > 0.2$ and $|\frac{\Delta|\mathbf{B}|}{|\mathbf{B}|_{mean}}| < 0.2$,

where $B_N$ is the field component perpendicular to the discontinuity plane. Different from *Smith* (1973), $\left|\frac{\Delta|B|}{|B|_{mean}}\right| > 0.2$ is not required for the TDs in this study. In the near-sun environment with quite low plasma beta, $\left|\frac{\Delta|B|}{|B|_{mean}}\right| > 0.2$ causes strong imbalance of the magnetic pressure across the TDs, difficult to be balanced by the plasma thermal pressure therein. Besides, the variation of magnetic field magnitude is a sufficient but unnecessary condition for TDs. Therefore, such a requirement may exclude many potential TD events near the sun and is removed in this study. By such classification criteria, we determine the types of these IDs and further investigate their characteristics statistically.

## 3. Results

Figure 1 presents the overview observation of PSP on Orbits 4 and 5, from December 1, 2019 to August 1, 2020. The heliocentric distance is shown in Figure 1a, displaying two perihelia at 0.13 au on January 29, 2020 and June 7, 2020 respectively. Figures 1b-1e exhibit the six-hour averaged solar wind parameters, including the radial ($B_R$) and tangential ($B_T$) components of the interplanetary magnetic field, the proton speed in radial direction ($V_R$), and proton density ($N_p$). The power-function fittings to the field data (the orange curves in Figures 1b-1c) yield $r^{-2.16}$ and $r^{-1.38}$ spatial scaling of $|B_R|$ and $|B_T|$, roughly consistent with the expectation for a spherically expanding magnetic field. The radial speed shows no clear spatial scaling but strong localized peaks and dips corresponding to fast and slow solar winds emerging from different regions on the sun surface. Around the perihelia the proton density reaches its maximum beyond 1000 cm$^{-3}$, at least one order of magnitude higher than the typical value measured at 1 au. During this period, we collect 3948 IDs in total, with the ID occurrence rate shown in Figure 1f. Apparently, the occurrence rate jumps from < 1 h$^{-1}$ (events per hour) at r>0.4 au, to >10 h$^{-1}$ when PSP approaches the perihelia.

Figures 2a-2b present the histograms of the ID event number and the PSP detection time, as functions of heliocentric distance. It should be noted that the detection time is the cumulative time of operation of the FIELDS instrument in each bin of distance, rather than the dwell time of the spacecraft since there are some data gaps. To remove the effect of PSP detection time on the detected event numbers, we calculate the ID occurrence rate, as shown in Figure 2c, by dividing the event

number by time. For all the IDs at [0.13, 0.9] au, we find a r$^{-2.00}$ spatial scaling of the ID occurrence rate. Subsequently, the ID occurrence rate is decomposed into the RD and TD occurrence rates according to the classification criteria we have already introduced. Of all 3948 IDs, 3068 IDs (~78%) are clearly classified, including 2511 RDs (64%) and 557 TDs (14%). The proportions of TDs and RDs estimated in this study are consistent with the results by *Horbury et al.* (2001), which indicate 57% RDs and 11% TDs, and some other work (Neugebauer et al. 1984; Mariani et al. 1983; Söding et al. 2001). The ratio RD:TD=2511:557≈4.51 is also close to the value RD:TD≈4.55 estimated by *Neugebauer & Giacalone* (2010), who combined five previous studies of 3806 IDs in total (Smith 1973; Neugebauer et al. 1984; Lepping & Behannon 1986; Söding et al. 2001; Horbury et al. 2001). Figures 2d-2e show the RD and TD occurrence rates. Interestingly, the RD occurrence rate has a steeper spatial scaling than the ID occurrence rate, exhibiting a power-law index of -2.17. In contrast, the decrease of TD occurrence rate is less steep. Figure 2f shows the relative occurrence rate $f_{RD}/f_{TD}$, where $f_{RD}$ and $f_{TD}$ represent the occurrence rates of RD and TD. $f_{RD}/f_{TD}$ is generally greater than 5 within 0.3 au, while at r>0.4 au $f_{RD}/f_{TD}$ fluctuates slightly between 0.6 and 1.7, exhibiting distinct evolution with distance.

From the identified TDs and RDs, their jump conditions can be revealed in more detail. Figures 3a-3b display the 2-D joint distributions of the field magnitude change $\frac{\Delta|\mathbf{B}|}{|\mathbf{B}|}$ and rotation angle <$\mathbf{B}_1$, $\mathbf{B}_2$> during the TD crossings, as a function of heliocentric distance. Here <$\mathbf{B}_1$, $\mathbf{B}_2$> represents the angle between the magnetic fields $\mathbf{B}_1$, $\mathbf{B}_2$ on two sides of the TD. The result shows that $\frac{\Delta|\mathbf{B}|}{|\mathbf{B}|}$ tends to vanish over the whole distance from 0.13 to 0.9 au, possibly due to the pressure equilibrium in the low plasma beta environment near the sun (Adhikari et al. 2020). However, within ~0.5au $\frac{\Delta|\mathbf{B}|}{|\mathbf{B}|}$ covers a wider range with its extrema larger/smaller than 1.5/-1.5, indicating the existence of dynamical structures possibly motivated by the intensely turbulent environment in the pristine solar wind (Greco et al. 2008, 2009; Servidio et al. 2011), which may be dissipated afterwards during its outward propagation. For a quantitative analysis of <$\mathbf{B}_1$, $\mathbf{B}_2$> evolution, an angular interval [α, β] is defined in each bin of heliocentric distance in such a way that 10% of the TDs in this bin have <$\mathbf{B}_1$, $\mathbf{B}_2$> smaller than α while another 10% TDs show <$\mathbf{B}_1$, $\mathbf{B}_2$> larger than β. In other words, the interval [α, β] covers the middle 80% TDs in each bin of distance. Two magenta curves in Figure 3b display such intervals at all distances. It can be seen that near the perihelion at 0.13 au, <$B_1$, $B_2$> of the TDs has a broader distribution centered around 45°. As the distance increases, <$B_1$, $B_2$> distribution gradually shrinks and is concentrated around 90°. Figures 3c-3d present the jump conditions across the RDs, including <$B_1$, $B_2$> and <$B_1$, $B_2$>$_{L-M}$ defined as the angle between the L-M components (i.e., the in-plane components) of the magnetic field on two sides of the RD. The magenta curves in Figure 3c have the same definition as that in Figure 3b. Different from TDs, the field lines during RD crossings rotate more regularly, exhibiting primarily an reversal of in-plane components (<$B_1$, $B_2$>$_{L-M}$→180°) with the total field rotating 30° (<$B_1$, $B_2$>→30°) near 0.13 au. At 0.3 au, such a regularity nearly vanishes. Furthermore, <$B_1$, $B_2$> of the RDs shows a clear upward trend as the distance increases, from ~30° at 0.13 au to ~70° at r>0.6 au.

Since <$B_1$, $B_2$> and <$B_1$, $B_2$>$_{L-M}$ are quite different for the near-sun and distant RDs, we deduce a special subgroup of RDs mainly existing within 0.3 au (see Figures 3c-3d). If the peaks of <$B_1$, $B_2$> at 30°, and <$B_1$, $B_2$>$_{L-M}$ at 180° near the sun are related to the same subgroup of RDs, these RDs must have the magnetic field with the dominant component in the direction perpendicular to the RD planes. For example, for a RD with <$B_1$, $B_2$>=30° and <$B_1$, $B_2$>$_{L-M}$=180°, it can be estimated as $\frac{|B_N|}{|B|} = \cos 15° \approx 0.966$. To verify this conjecture, we plot the 2-D RD event distribution in Figure 4, as a function of <$B_1$, $B_2$>$_{L-M}$ and <$B_1$, $B_2$>. The RDs at r<0.2 au (1469 events) and r>0.4 au (186 events) are displayed by the green and red dots respectively, while the RDs in the transition region [0.2, 0.4] au are excluded in order to make the distinction between the near-sun and distant RDs more noticeable. The dashed line is the boundary of <$B_1$, $B_2$>$_{L-M}$=<$B_1$, $B_2$>. Indeed, the RD event distribution at r<0.2 au can be regarded as the combination of a clustered component in the area of small <$B_1$, $B_2$> and large <$B_1$, $B_2$>$_{L-M}$ as marked by the ellipse, and a randomly distributed background (the green dots outside the ellipse). For the RDs at r>0.4 au, their distribution looks similar to the background component of the RDs at r<0.2 au. Thus, a special subgroup of RDs (see the green dots clustered in ellipse), characterized by small <$B_1$, $B_2$> which also implies quasi-parallel or antiparallel propagation of the RDs to the background field, do exist in the near-sun solar wind. However, this type of RDs almost disappear at r>0.3 au (see Figures 3c-3d), and therefore have not been revealed before the advent of PSP. Their limited occurrence region and similar field deflections may indicate their origin from the sun, or a certain formation mechanism at work in the pristine solar wind where the plasma is still experiencing acceleration. On the other hand, another subgroup of RDs observed with random field deflections (see red dots and green dots outside the ellipse), exist and propagate more extensively in the solar wind which are detected at all distances between 0.13 and 0.9 au in this study.

Figure 5 presents the joint distributions of the discontinuity orientation versus heliocentric distance, where <**N**, **R**> and <**N**, **B**> are defined as the angle between the normal of the discontinuity plane **N** and the radial direction **R**, and the background magnetic field **B**, respectively. The magenta curves in Figure 5 show the medians of <**N**, **R**> or <**N**, **B**> as functions of distance. In the near-sun region within 0.25 au, the TDs tend to slope towards the radial direction, as indicated by <**N**, **R**> being close to 90°, while the RDs are likely to face the radial direction with <**N**, **R**> mainly smaller than 45°. As the distance increases, median <**N**, **R**> decreases for the TD from ~70° to ~30°, but slightly increases for the RD from ~30° to ~45°. In contrast, <**N**, **B**> of RDs shows quite distinct characteristics and evolutions. The RDs have a parallel propagation tendency within 0.3 au, while beyond the distance no obvious trend is found in RD orientation with respect to the magnetic field.

In order to estimate the characteristic thicknesses of the TDs and RDs in space, we employ a comprehensive analysis by considering the proton data from SPC, as described below. First, for each ID we define a 16-second window centred at the ID, and average the proton data in this window as the environment parameters for the corresponding ID, including the background solar wind velocity $\mathbf{V}_{sw}$ and plasma density, if the proton data are available in this window. The ID events without proton data or with "unreliable proton data" marked by the quality flags of SPC in the windows are removed. After the selection, 296 TDs and 1727 RDs, which are mostly located within 0.6 au, are reserved for thickness estimation. Then, we define the ID duration as the double length of the period when $\frac{\partial B_L}{\partial t} > \frac{1}{3} \max\{\frac{\partial B_L}{\partial t}\}$, where $\frac{\partial B_L}{\partial t}$ is the time derivative of the field maximum-variation component determined by MVA, and $\max\{\frac{\partial B_L}{\partial t}\}$ is the maximum of $\frac{\partial B_L}{\partial t}$ generally measured at the center of the discontinuities. Since the TDs are stationary in the plasma rest frame, their normal velocities relative to the PSP spacecraft can be obtained as $(\mathbf{V}_{TD,PSP})_N = (\mathbf{V}_{sw} - \mathbf{V}_{PSP})_N$, where $\mathbf{V}_{sw}$ and $\mathbf{V}_{PSP}$ are the velocities of the solar wind and the PSP spacecraft, and the notation $()_N$ indicates the normal component of the vector in brackets to the ID plane. The TD thicknesses are therefore directly calculated as the products of the event duration and $(\mathbf{V}_{TD,PSP})_N$. To estimate the RD thicknesses, we need to determine their propagation velocities first, since the RDs are propagating structures in the plasma rest frame. According to the Walén relation, the normal flow speed in the RD rest frame correlates with the normal magnetic field, as $|V_{N,RD}| = \frac{|B_N|}{\sqrt{\mu_0 \rho}}$ (Sonnerup et al. 1987), while its direction (parallel or antiparallel to the N direction) is still unknown. In the RD rest frame, the tangential electric field carried by the plasmas should be continuous during the crossing since the discontinuity is stationary and thus the magnetic field is time-independent. Hence, we bring both possibilities—$|V_{N,RD}|$ and -$|V_{N,RD}|$—into the ideal Ohm's law $\mathbf{E} = -\mathbf{V} \times \mathbf{B}$, and check which one produces a continuous tangential electric field in the RD rest frame. So far, both the sign and the magnitude of $V_{N,RD}$ have been determined. Obviously, the RD propagation velocity in the plasma rest frame is $-V_{N,RD}$. Since the relative movement of the solar wind to PSP spacecraft can be similarly calculated by $\mathbf{V}_{sw} - \mathbf{V}_{PSP}$, the RD velocity relative to spacecraft can be determined by $(\mathbf{V}_{RD,PSP})_N = (\mathbf{V}_{sw} - \mathbf{V}_{PSP})_N - V_{N,RD}$. The RD thicknesses are then estimated by multiplying the event duration and $(\mathbf{V}_{RD,PSP})_N$ together. A detailed demonstration of this process can be found in a companion paper focusing on the case study of IDs at ~0.13 au (Liu et al. 2021).

Figures 6a-6b exhibit the joint distributions of the TD and RD thickness normalized by the ion inertial length ($d_i$). The normalized TD thickness shows neither a typical value nor significant spatial scaling. We can only resolve a rough range of the TD thickness distribution as [5, 35] $d_i$ (see two horizontal dashed lines in Figure 6a) at [0.13, 0.8] au, based on 296 TDs detected here. In contrast, the RD thickness displays significant characteristics. The crosses in Figure 6b show the mean values of the RD thicknesses in each bin of heliocentric distance, while the black curve exhibits the power-function fitting model. First, the RD thicknesses at the same distance are comparable, exhibiting a relatively narrow distribution. Moreover, the normalized RD thickness has $r^{-1.09}$ scaling over the distance from 0.13 to 0.6 au. Considering that $d_i$ is inversely proportional to the square root of plasma density and thus proportional to the heliocentric distance, the RD thickness in units of kilometers, in fact, does not change a lot with heliocentric distance. The average thickness of 1727 RDs is estimated to be 574 km. Similarly, the TDs actually become thicker as the distance increases.

Figure 6c shows the joint distribution of the RD propagation velocity in the plasma rest frame. Positive and negative velocities indicate the outward (anti-sunward) and inward (sunward) propagation, respectively. As can be seen, the outward propagating RDs predominate among all RDs. A minority of RDs also exhibit inward propagation, implying potential mechanisms, such as turbulence, at work locally generating and pitching RDs in all directions. Then we focus on the outward propagating RDs and calculate their mean propagation velocity as a function of distance, as shown by the plus signs in Figure 6d. The fitting shows $r^{-1.03}$ scaling of the RD propagation velocity, consistent with the expected $r^{-1}$ scaling from Walén relation. It implies that the RDs at smaller distances always move faster than those at larger distances, which may in turn influence the particles in the solar wind through some mechanisms like Fermi process (Guo et al. 2014; Park et al. 2015; Liu et al. 2019).

## 4. Discussion and Conclusions

In this study, we investigated the interplanetary discontinuities at [0.13, 0.9]au, by utilizing the field and proton data from PSP on Orbits 4 and 5 from December 1, 2019 to August 1, 2020. A total of 3948 events were collected, of which 3068 events were clearly identified as the RDs or TDs. We further investigated the RD and TD characteristics, including their occurrence rates, the jump conditions, the orientations, the thicknesses, and the propagations in the plasma rest frame. The main conclusions are summarized as follows:

1) IDs are more abundant in the innermost region of the heliosphere. The ID occurrence rate has $r^{-2.00}$ spatial scaling at [0.13, 0.9] au. The inhomogeneity of the RD spatial distribution is more obvious than that of TD.

2) The relative occurrence rate $f_{RD}/f_{TD}$ decreases with the heliocentric distance, from $f_{RD}/f_{TD} \approx 8$ at r<0.3 au to $f_{RD}/f_{TD} \approx 1$ at r>0.4 au.

3) The magnetic field tends to retain a constant magnitude across the TDs at [0.13, 0.9] au. In the near-sun region, the field rotates statistically 45° across the TDs, and 30° with opposite tangential component across the RDs. Such regularities disappear beyond 0.3 au.

4) Within 0.25 au, the normal direction of the discontinuity plane tends to be perpendicular to the radial direction for the TDs, and parallel for the RDs. The RDs which propagate parallel or antiparallel to the field lines predominate within 0.3 au.

5) The TD thickness normalized by the ion inertial lengths shows no clear spatial scaling and is mostly distributed in the range of [5, 35] $d_i$. The RD thickness in units of $d_i$ decreases with the distance, following $r^{-1.09}$ spatial scaling.

6) The RDs of the outward (anti-sunward) propagation in the plasma rest frame predominate, with their propagation speeds proportional to $r^{-1.03}$.

Among all the issues regarding the IDs, the proportions of different types of discontinuities and their occurrence rates have been mostly investigated but also intensely debated. *Mariani et al* (1983) found by searching the data from Helios 1 and 2 about twice as many RDs as TDs over the heliocentric distance [0.3, 1] au. *Smith* (1973) and *Neugebauer et al.* (1984) reported ratios as RD:TD=44:18 and RD:TD=117:19 respectively. *Lepping* and *Behannon* (1986) showed that the RD to TD ratio decreases with heliocentric distance, being 1.16, 0.83, and 0.67 at the distances 0.46, 0.72, and 1.0 au. Such large discrepancies between the estimation of RD to TD ratio have been mainly attributed to different identification criteria in previous studies. According to this study, this ratio is proved to clearly depend on the heliocentric distance. From 0.13 to 0.9 au, the RD to TD ratio could decrease by nearly an order of magnitude. This may be the main factor responsible for the different estimation of RD to TD ratio in previous studies. On the other hand, the decrease of RD to TD ratio mainly occurs within 0.4 au and becomes less obvious beyond this distance.

Regarding the ID occurrence rate, it has been widely accepted that it depends on the heliocentric distance. Specifically, several studies have reported the spatial scaling of the occurrence rate $f_{ID}$ in different forms, including $f_{ID} \propto r^{-\alpha}, \alpha \in [-1.3, -0.78]$ (Lepping & Behannon 1986; Söding et al. 2001), $f_{ID} \propto e^{-(r-1)/5}$ (Tsurutani et al. 1996), and $\Delta f_{ID} = -13\Delta r$ (Mariani et al. 1973), based on the measurements in different regions over [0.3, 19] au. Most researchers attribute the occurrence rate decrease to a combination of geometric effect due to the radial wind expansion and the discontinuity thickening which breaks the selection criteria (Lepping & Behannon 1986; Mariani et al. 1973), while some others suggest that ID annihilation should exist (Söding et al. 2001; Tsurutani & Smith 1979). The PSP results bring new insight into this issue. Figures 2d-2e reveal much sharper decrease of $f_{RD}$ than $f_{TD}$ with the distance. Neither the geometric effect nor the discontinuity thickening can explain the phenomenon, since the geometric effect changes $f_{RD}$ and $f_{TD}$ equally and the discontinuity thickening happens to TDs rather than RDs according to our results. Consequently, we infer that there must be a decay channel for RDs within 0.4 au to be responsible for the sharper $f_{RD}$ scaling.

**Acknowledgement.**

We acknowledge the *Parker Solar Probe instrument* teams for the years of work and the convenient data access. The SWEAP and FIELDS investigation and this publication are supported by the PSP mission under NASA contract NNN06AA01C. PSP data are available to the public via NASA's Space Physics Data Facility (SPDF) at https://spdf.gsfc.nasa.gov/pub/data/psp/. We also acknowledge N. E. Papitashvili for providing the continuous PSP trajectory data which are publicly accessible at https://spdf.gsfc.nasa.gov/pub/data/psp/ephemeris/helio1day/. This work was supported by NSFC grants 41821003 and 41874188.

**Figure captions**

**Figure 1.** PSP observations on Orbits 4 and 5. From top to bottom, the panels show the spacecraft distance from the sun, the magnitudes of the radial and tangential components of the magnetic field ($|B_R|$ & $|B_T|$), the proton speed and number density, and the occurrence rate of the interplanetary discontinuities. The vertical dotted line indicates the aphelion on April 3, 2020 which divides Orbits 4 and 5. The oranges curves in panels (b-c) are the power-function fittings of $|B_R|$ and $|B_T|$, with the coefficient of determination $R^2$ being 0.904 for $|B_R|$ and 0.788 for $|B_T|$. The uncertainties of the power-law indices correspond to 95% confidence bounds.

**Figure 2.** Histograms as a function of heliocentric distance of (a) the number of events, (b) the PSP detection time, and (c-e) the occurrence rates of the total IDs, RDs and TDs. Panel (f) shows the ratio of RD occurrence rate to TD occurrence rate. The oranges curves in panels (c-d) are the power-function fittings of ID and RD occurrence rates, with the coefficient of determination $R^2$ being 0.983 and 0.985 respectively. The uncertainties of the power-law indices correspond to 95% confidence bounds.

**Figure 3.** Joint distributions of the event number as a function of heliocentric distance and (a-b) the magnitude change or the rotation angle of the magnetic field crossing the TDs, and (c-d) the rotation angles of the total or in-plane magnetic field crossing the RDs. The magenta curves in panels (b-c) indicate the angular intervals which covers the middle 80% events in each bin of distance.

**Figure 4.** The 2-D event distribution as a function of the rotation angle of total magnetic field and the rotation angle of the in-plane magnetic field. Green and red dots present the RDs at r<0.2 au and r>0.4 au respectively. The dashed line presents the boundary of $<B_1, B_2>_{L-M}=<B_1, B_2>$, while the green ellipse corresponds to the RDs clustered in the area of small $<B_1, B_2>$ and large $<B_1, B_2>_{L-M}$ within 0.2 au.

**Figure 5.** Joint distributions of the discontinuity orientation, $<N, R>$ and $<N, B>$ vs. heliocentric distance. The magenta curves display the medians of $<N, R>$ or $<N, B>$ in each bin of distance.

**Figure 6.** Joint distributions of the event number as a function of heliocentric distance and (a-b) the thickness of TDs or RDs, and (c) the propagation velocity of RDs in the plasma rest frame. Panel (d) shows the mean RD propagation velocity as a function of heliocentric distance (the blue plus signs). In panel (b) the black crosses show the mean values of the RD thicknesses in each bin of distance, while the black curve is the power-function fitting of them with the coefficient of determination $R^2$ being 0.868. In panel (d), the red curve displays the power-function fitting of the mean RD propagation velocity with the coefficient of determination $R^2$ being 0.964. The uncertainties of the power-law indices correspond to 95% confidence bounds.

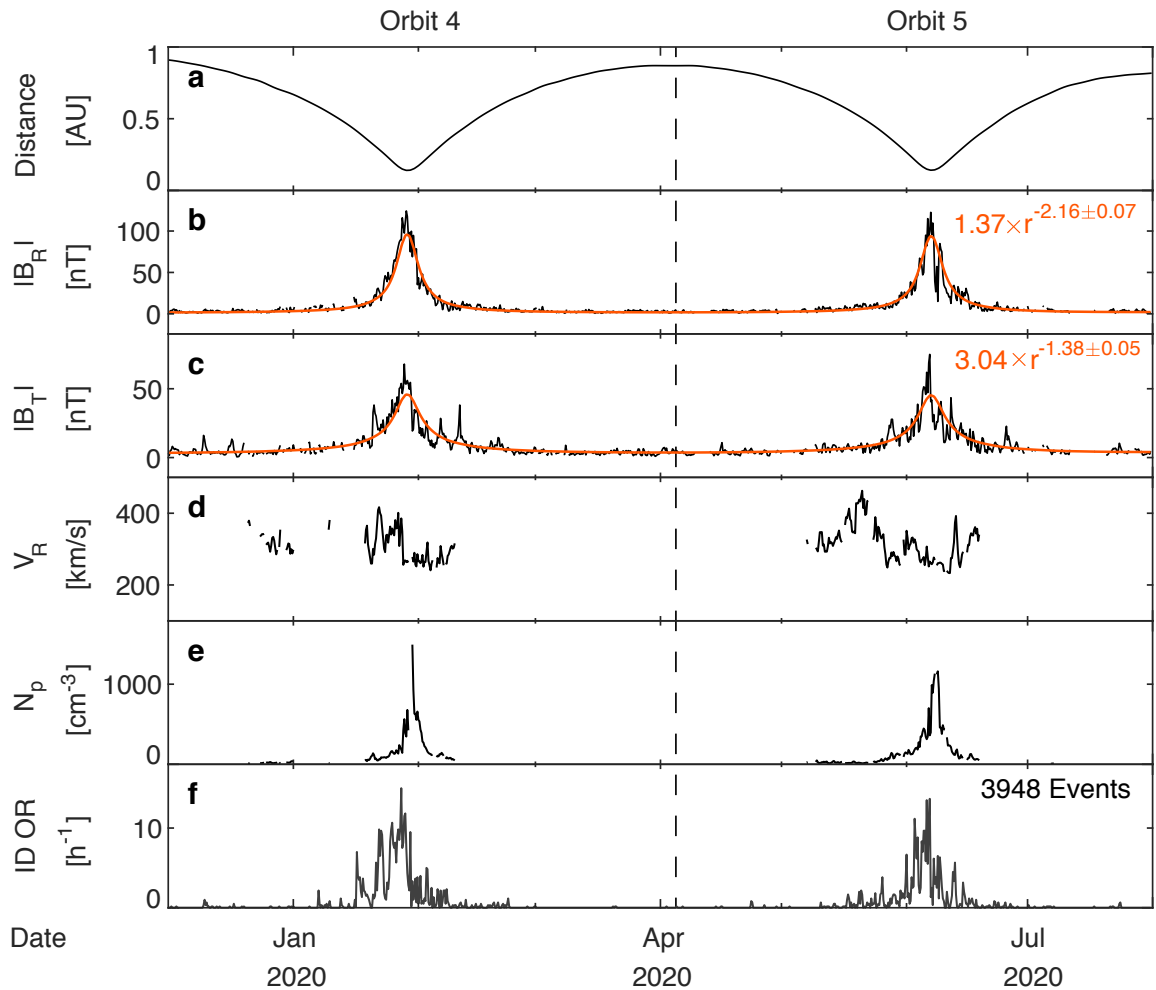

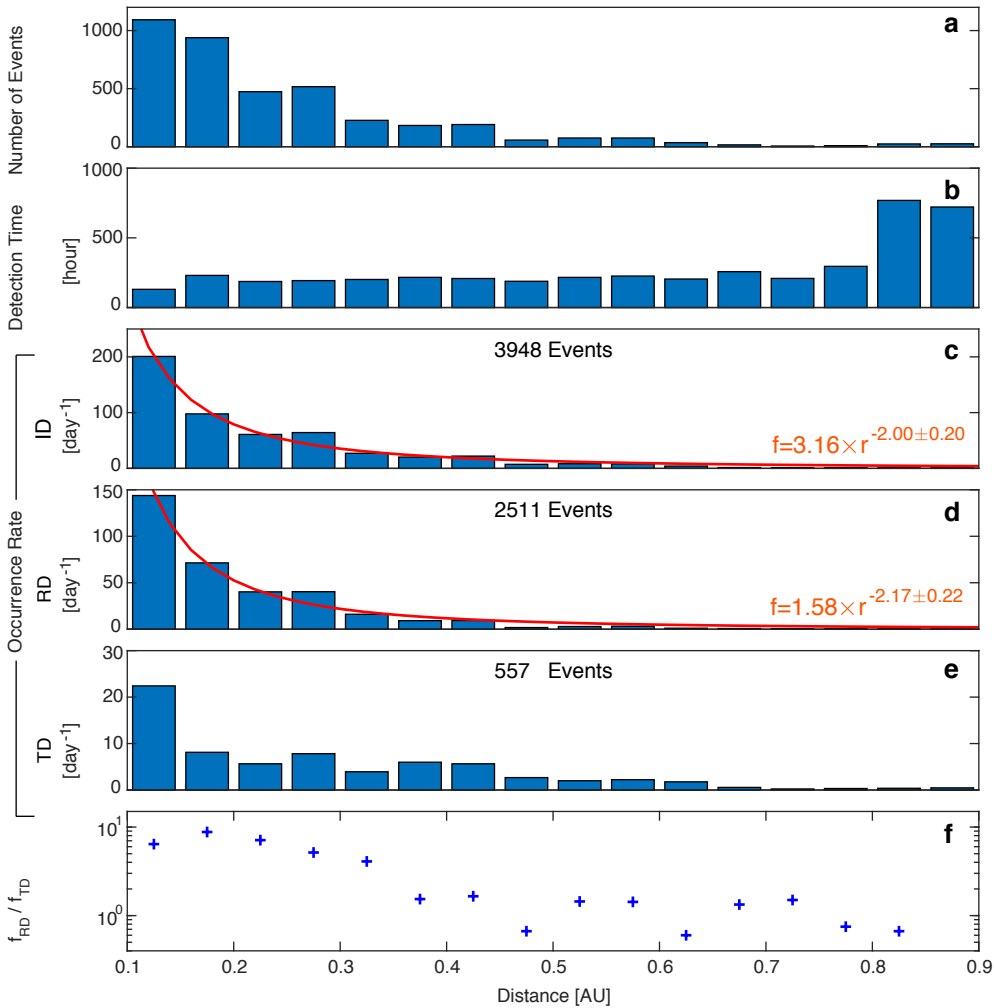

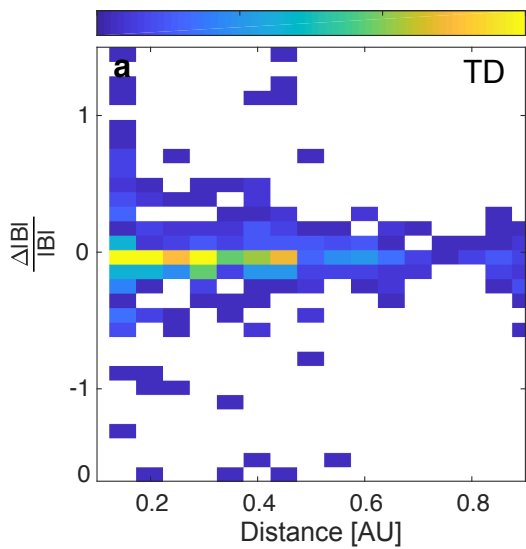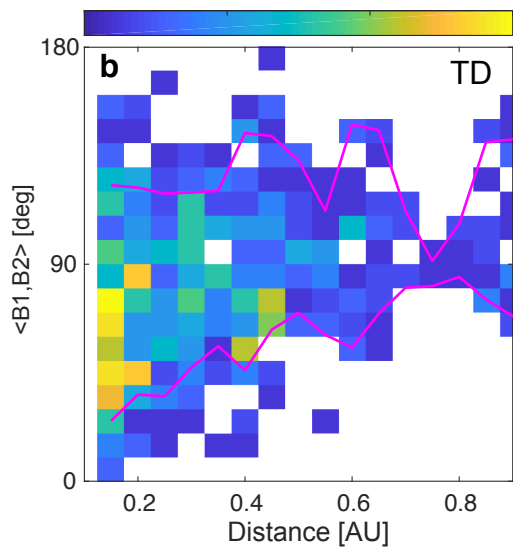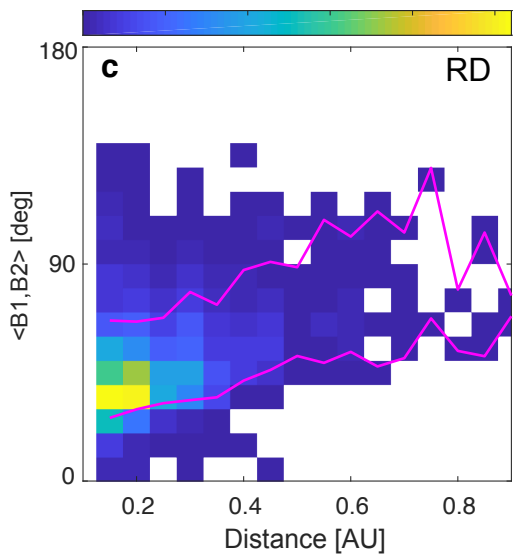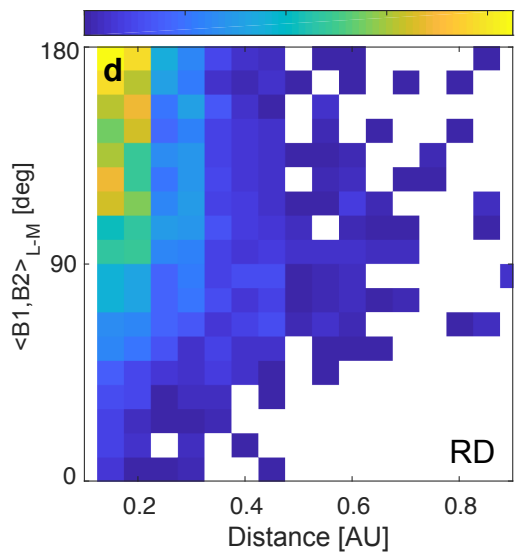

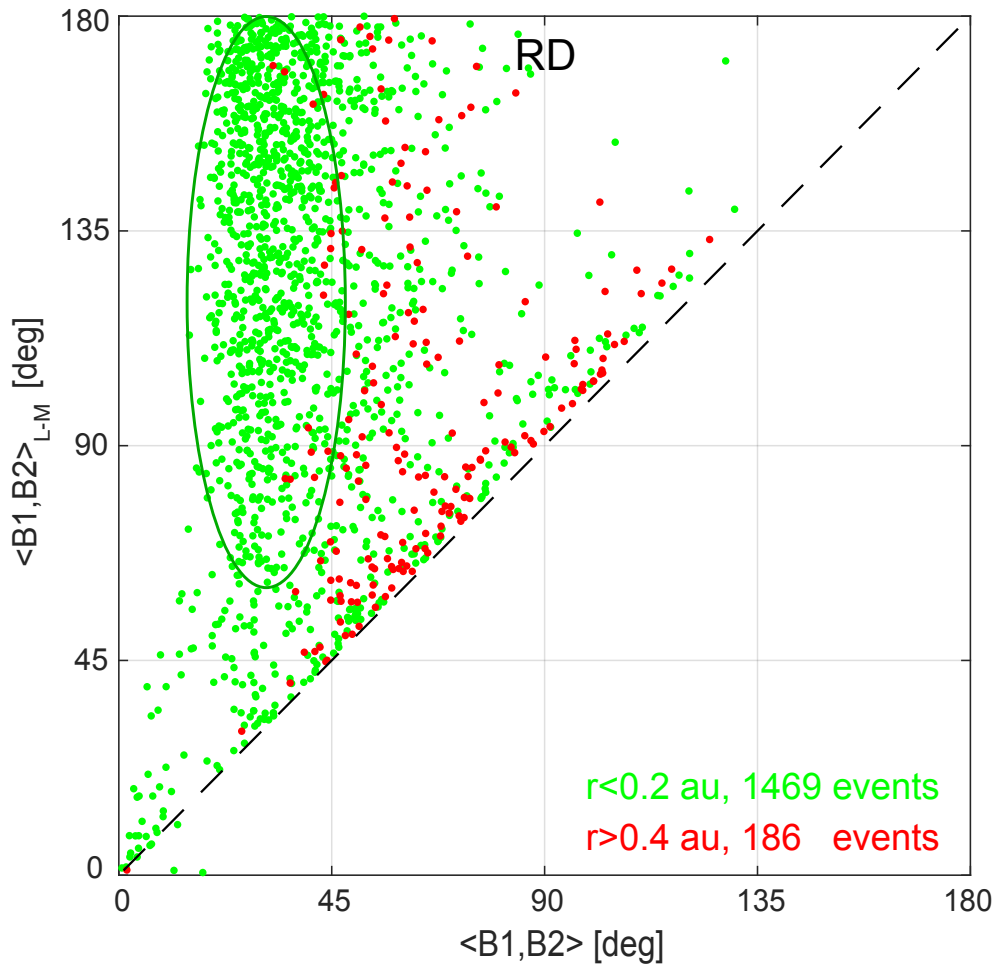

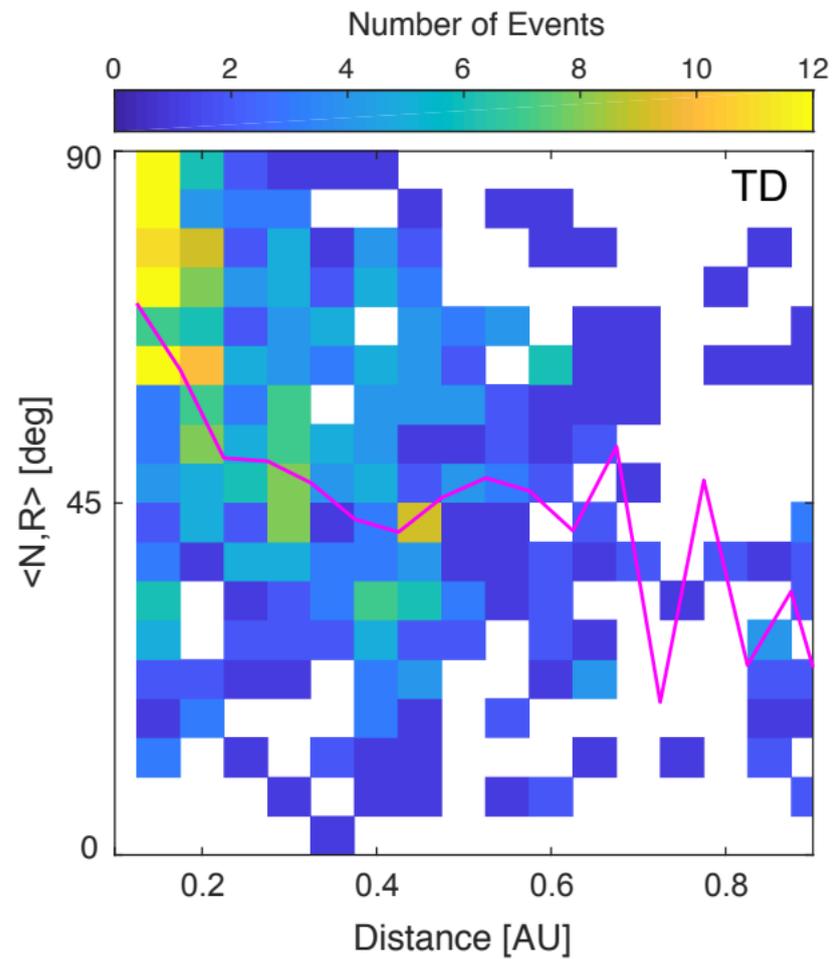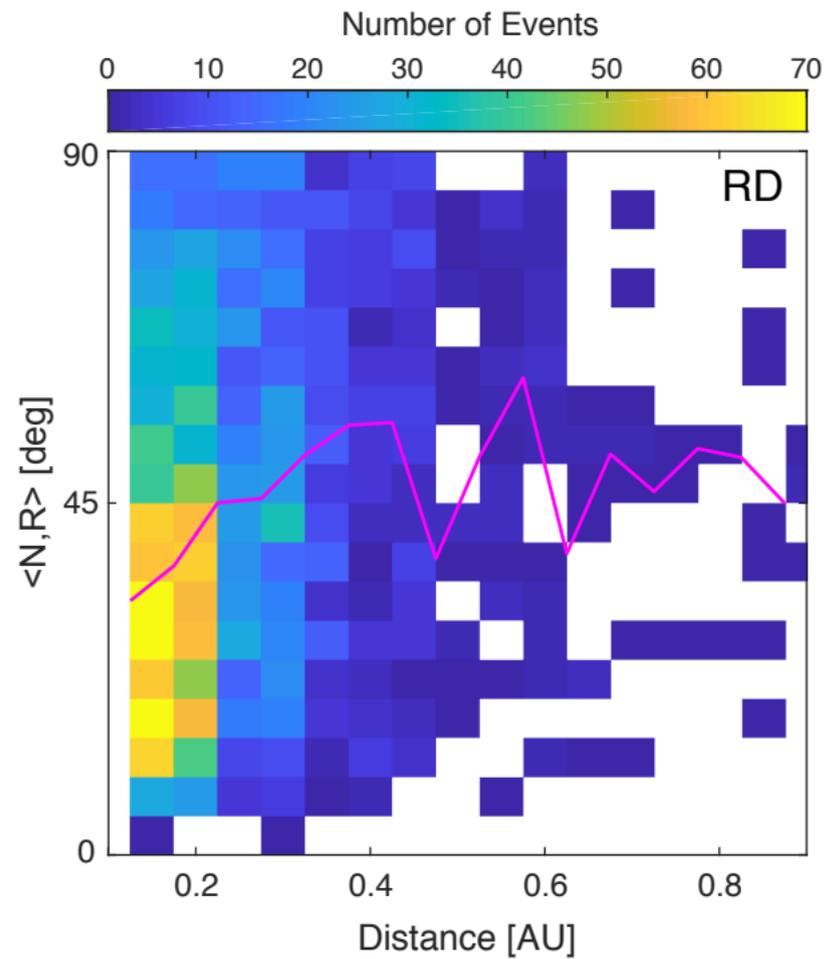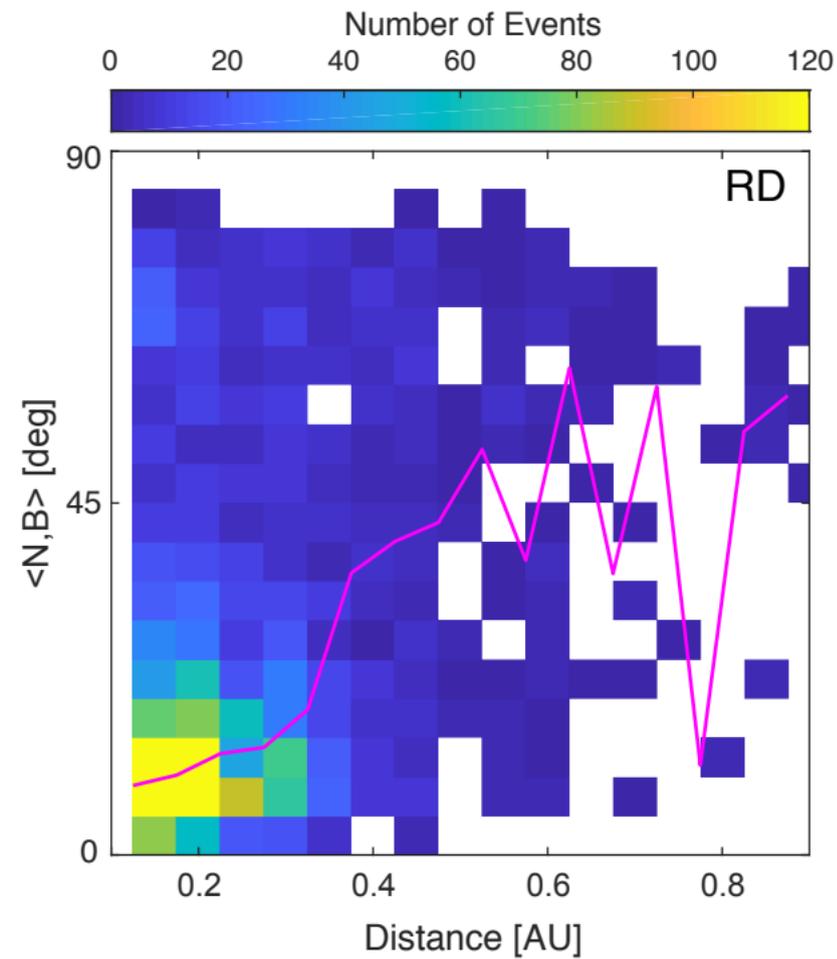

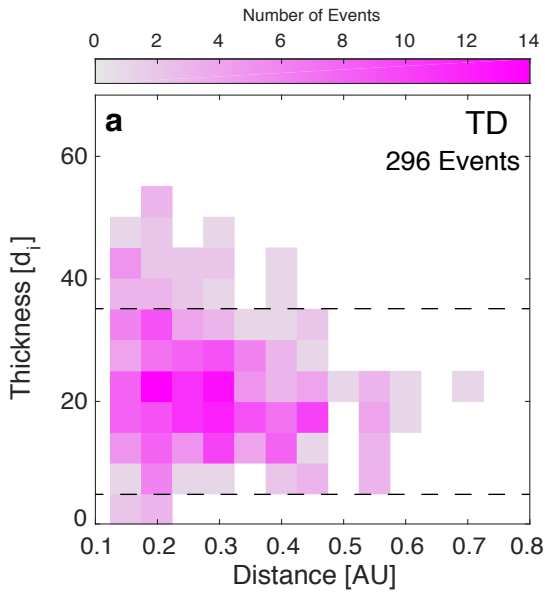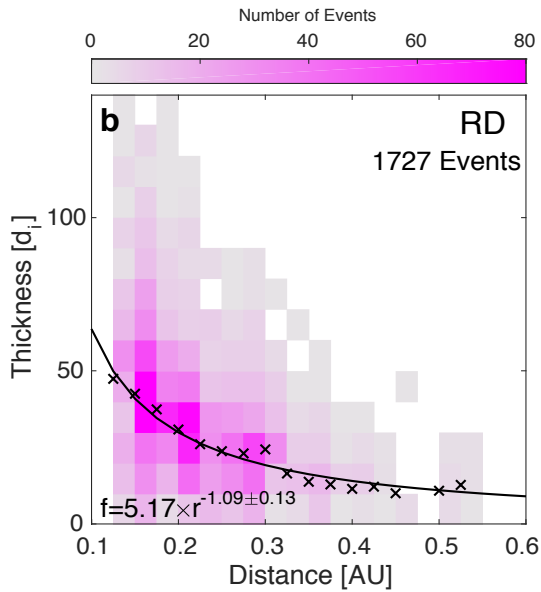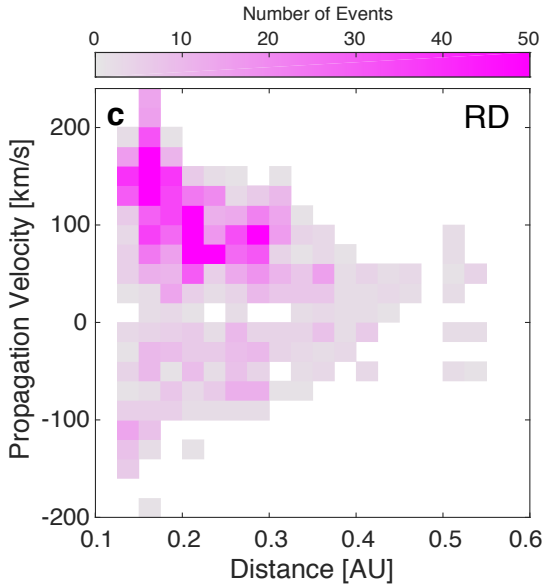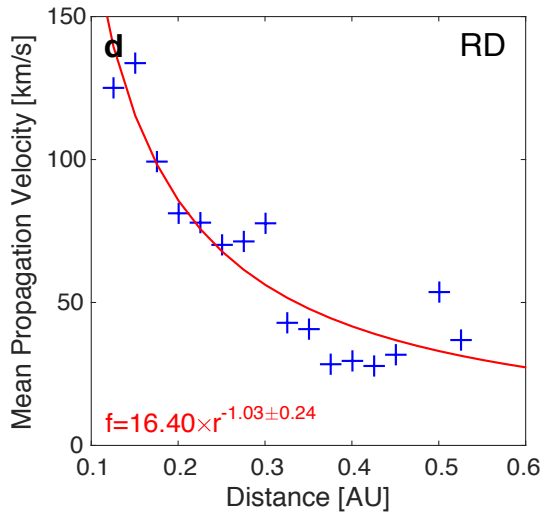